\documentclass[12pt,a4paper]{article}

\newcommand{\Z}{\ensuremath{{\mathsf{Z\!\!Z}}}}

\begin{document}

\title{(De)Localization in the Prime Schr\"odinger Operator\vspace{2cm}}
 \author{C\'esar R. de Oliveira {\small and} Giancarlo Q. Pellegrino\\
\vspace{-0.6cm}
\small
\it Departamento de Matem\'{a}tica -- UFSCar, \small \it S\~{a}o Carlos, SP,
13560-970 Brazil\\}
\date{}

\maketitle

\vspace{1cm}
\begin{abstract} It is reported a combined numerical approach to study the
localization properties of the one-dimensional tight-binding model with
potential modulated along the prime numbers. A localization-delocalization
transition was found as function of the potential intensity; it is also
argued that there are delocalized states for any value of the potential
intensity.\\

\end{abstract}

\clearpage

Problems related to the distribution of prime numbers are usually
interesting and difficult. Maybe, the most challenging open problem in
mathematics is the so-called Riemann's hypothesis on the distribution of
the (nontrivial) zeros of his zeta function, a problem directly related
to the distribution of prime numbers and other mathematical issues. Also,
the interplay between number theory, with the set of prime numbers
playing a major role, and physics has been fruitful; see the short note
by Wolf~\cite{W} and references therein for recent discussions.

In the last years there has been a large interest in spectral properties of
discrete Schr\"odinger operators on~$l^2(\Z)$
\[ (H_{V}u)_n = u_{n+1}+u_{n-1} + \lambda V_n u_n,
\] with nonrandom finite valued potentials~$V=\{V_n\}$ (see
\cite{AG,H,D,deOP,ZM,deOL} and references therein); $\lambda>0$ is the
potential intensity. Such one-dimensional models with potentials along
almost periodic sequences taking a finite number of values~\cite{AG,H,D},
like substitution and circle sequences, have been dominated by singular
continuous spectra. Recall, however, that for the cases of periodic and
random potentials $V$ it is well known that the hamiltonian~$H_V$ has
absolutely continuous and point spectra, respectively~\cite{CL}; these
are mathematical characterizations of delocalized and localized
(insulator) behaviours, respectively.

There has also been recent work on Schr\"odinger operators with sparse
potentials~\cite{SS,R,KLS,CES,HK}. A characteristic defining a sparse
potential is the constancy of its values in a sequence of increasing
intervals. In this case one can have point, absolutely continuous and
singular continuous spectra, depending on the growing speed of the gaps and
how the intensity of the bumps behaves at infinity. There is, in fact, some
competition among different properties of the potential, resulting in rich
spectral possibilities.

The prime numbers are straight related to both kind of potentials discussed
above, namely nonrandom and sparse. It is well known that the set of prime
numbers is not finite and that there are arbitrarily large gaps between
consecutive primes, a property characterizing sparseness. By defining a
potential~$V_0=0$, $V_n=1$ if~$n+1$ is a prime number and~$V_n=0$ if not,
and extending symmetrically to negative values of~$n$, i.e., $V_{-n}=V_n$,
one gets an instance of finite valued sparse potential. The nonnegative
values of the potential~$V$ are $01101010001010001\cdots$.

The main aim of this Letter is to discuss numerically the spectral
properties of the corresponding discrete Schr\"odinger operator, which
will be shortly referred to as ``the prime Schr\"odinger operator.'' The
rigorous analytical study of this system, although desirable, seems to be
subtle and far from trivial. Since in this Letter we consider this problem
from just a numerical point of view, we find more appropriate to avoid the
mathematical spectral terms, and instead shall try to use the
physical ones localization (insulator) and delocalization to
characterize our results.

By exploring dynamical as well as geometrical quantities related to the
spectral properties of the model, we find a (rather smooth) transition from
predominantly delocalized states to localized ones, as the potential
intensity is increased. Nevertheless, some delocalized states always remain,
no matter how large is the potential. Also, the combined use of geometrical
and dynamical tools, supportive and to some extent complementary to each
other, seems to be applicable to other cases with coexisting spectral
components.

Here two different sets of numerical tools are used in such investigations.
The {\em dynamical tools} are the average probability for the system to
return to its initial state~$u_0=\delta_{n,n_0}$, concentrated on the
site~$n_0$,
\[ C(t) = \frac{1}{t}\int_0^t |\langle u(s)|u_0\rangle|^2 ds,
\] and the  second moment
\[ d_{2}(t) =  \sum_{n} (n-n_0)^2|u_n(t)|^2,
\]
$u=(u_n)_{n\in\Z}$. Recall  that for large values of~$t$ it is expected that
$C(t)\sim t^{-D_2}$ and
$d_{2}(t)\sim  t^{2\beta}$, with $D_2$ being the correlation dimension of
the spectral measure  associated with the initial state~\cite{G,BCM,L}.
Continuous spectra are usually numerically characterized by $0<D_2,\beta$;
although it is possible to have exceptional systems with  point  spectrum
and with any value $0\le \beta<1$~\cite{dRJLS}.

Due to  the  rather irregular distribution of prime numbers~\cite{W} we
suspect the quantum  dynamics  generated by~$H_V$ can strongly depend on the
initial wave function, at least for small times; therefore,  the  initial
condition was always concentrated on $n_0=0$. These dynamical tools were
restricted to $0.5\le \lambda$, since for  small~$\lambda$ it is not
possible to neglect logarithmic corrections to the algebraic behaviours of
$C(t)$ and $d_2(t)$~\cite{ZM}.

The other kind of tools, {\em geometrical tools}, are the direct inspection
of eigenfunctions and the {\em Lyapunov exponent} (LE)~$\gamma$, also called
{\em inverse localization length}, calculated for each eigenfunction of the
finite basis approximation. The typical basis size used here is~$10^4$; the
robustness of the numerical calculations with respect to the basis size was
verified in each case. Localization, i.e., point spectrum, is characterized
by nonvanishing~$\gamma$.

From the Thouless formula~\cite{CL,CFKS}, in particular from its proof,
one sees that the LE at energy~$E$ can be estimated as
\[
\gamma(E)\approx \frac{1}{N}\sum_{j=1}^{N}\ln|E-E_j|,
\] with~$E_j$ running over all eigenvalues (distinct from~$E$) of the finite
basis approximation of size~$N$.

In Fig.~1 it is shown the Lyapunov exponent~$\gamma$ as function of the
energy for some values of the potential intensity~$\lambda$.
For~$\lambda=0.5$ the vanishing of the LE for most energies~$E$ indicates
the predominancy of delocalized states; by increasing the potential
intensity the LE becomes different from zero, except in a left
neighborhood of~$E=0$. This phenomenon was verified for all values of
potential intensity considered (up to~$\lambda=40.0$), i.e.,  the LE
vanishes just below~$E=0$. This strongly indicates a transition from
delocalization to localization with a remaining set of delocalized states
near~$E=0$.

The above conclusions were also supported by numerical solutions of the
Schr\"odinger equation and computation of dynamical quantities (also with
lattice size up to~$10^4$). In Fig.~2 it is shown~$d_{2}(t)$ for two values
of~$\lambda$; $d_{2}(t)$ saturates for~$\lambda=6.0$ (as well as~$C(t)$;
not shown), indicating the presence of localized states; while
for~$\lambda=0.5$ the values of~$d_{2}(t)$ increase until a critical
probability density value is reached at the basis border ($ |u|^2 <
10^{-6}$ was used), indicating the presence of delocalized states. Notice
that, due to the presence of both localized and delocalized states for
the prime Schr\"odinger operator, it can be difficult to retrieve
information on its spectral type only from dynamical quantities. This
justifies the choices of potential intensity values in~Fig.~2, used to
exemplify the spectral transition: for~$\lambda=0.5$ most states are
delocalized, while for~$\lambda=6.0$ the energy band corresponding
to delocalized states is very narrow.

Of course a numerical vanishing LE is not exactly zero. The LE~$\gamma_{\rm
p}$ for a periodic case was computed with the above method; since it should
be zero in the spectrum of the corresponding Schr\"odinger operator,  any
$\gamma\le \gamma_{\rm p}$ was in practice considered to indicate the
presence of delocalized states for the prime case. This was used to estimate
the length of the interval~$[-b(\lambda),0]$ of the ``delocalized band'' as
function of~$\lambda$. As another check for such transition, the dynamical
quantities~$d_{2}(t)$ and~$C(t)$ were computed for short times taking into
account two different sets of eigenvectors: those whose eigenvalues were in
the delocalized band~$[-b(\lambda),0]$, and those in intervals with
positive~LE. The distinctions in their qualitative behaviours are evident.
For example, in Fig.~3 the value $\lambda=1.0$ was fixed and~$d_{2}$ was
computed by restricting the dynamics to eigenvalues in the
range~$[-0.6,0.0]$ (a subset of~$[-b(1),0]$) and also to eigenvalues in
the range~$[1.6,2.0]$ (corresponding to positive~LE). In the former
case~$d_{2}(t)$ grows with time, while in the latter case it is clearly
bounded; such results also support the presence of mobility edges in
the spectrum of the prime Schr\"odinger operator.

Fig.~4 shows some values of~$b(\lambda)$; the best fitting line is also
shown and its slope is equal to~$-1$. Therefore, it is numerically found
that $b(\lambda)\sim \lambda^{-1}$, at least for large $\lambda$, so that
it does not vanish and the spectrum of the prime Schr\"odinger operator
should have a set of delocalized states for any potential intensity.
Notice that the eigenfunctions corresponding to eigenvalues in the
range~$[-b(\lambda),0]$ are extended over the finite bases, while they
are exponentially localized if the corresponding LE is greater than zero
(not shown here).

In summary, a delocalization-localization transition, with mobility edges,
was found numerically for the one-dimensional prime Schr\"odinger
operator, a transition as the intensity of the potential is increased;
however, there remains a band of delocalized states whose length scales
as~$\sim\lambda^{-1}$, assuring the presence of delocalized states for
any value of~$\lambda$. These results, and the calculations presented,
suggest that the same approach could be applied to cases with other
aperiodic potentials, for which one suspects to have simultaneously
continuous and point spectral components.

\vspace{1.0 cm} \subsubsection*{Acknowledgments} {\small We thank CAPES-DAAD
for supporting our visit (PROBRAL Project) to Prof.\ M.\ Schreiber's group
at Technische Universit\"at Chemnitz, where part of this work  was done.
We also thank Prof.\ W.\ F.\ Wreszinski for his patient cooperation. GQP
was supported by FAPESP and CRO was partially supported by CNPq.}

\clearpage \subsubsection*{Figure Captions}
\vspace{1cm}

{\bf Figure 1.} Lyapunov exponent~$\gamma$ versus energy for the prime
Schr\"odinger operator. a)~$\lambda=0.5$;  b)~$\lambda=1.0$;
c)~$\lambda=3.0$; d)~$\lambda=6.0$.

\vspace{2cm}

\noindent {\bf Figure 2.} The second moment~$d_{2}$ as function of time for
a)~$\lambda=0.5$ and b)~$\lambda=6.0$. Notice the different scales
(for~$\lambda=6.0$ the time evolution did not reach the basis border).

\vspace{2cm}

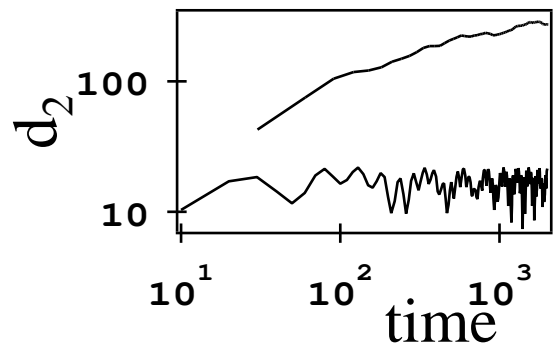
\noindent {\bf Figure 3.} The second moment~$d_{2}$ as function of time
for~$\lambda=1.0$ (base 10 $\log-\log$ scale). The upper curve was obtained
taking into account only the eigenvectors whose eigenvalues were in the
range~$[-0.6,0.0]$. Similar for the lower curve, but with eigenvalues in the
range~$[1.6,2.0]$. See also Fig.~1b.

\vspace{2cm}

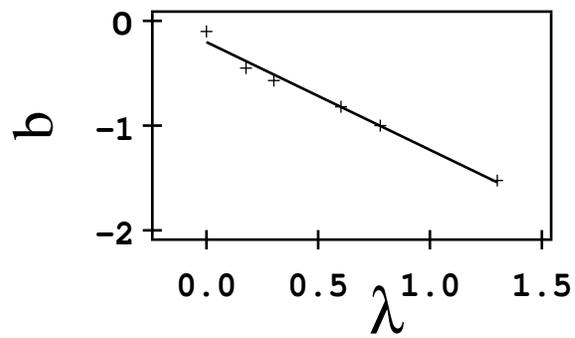
\noindent {\bf Figure 4.} Delocalized band length~$b$ as function of the
potential intensity~$\lambda$ (base 10 $\log-\log$ scale). The best fitting
line is also shown.

\end{document}